\documentclass[useAMS,usegraphicx,usenatbib]{mn2e}

\def\apj{ApJ}
\def\mnras{MNRAS}

\def\lt{${\rm L_{SX,tot}}\,\,$}
\def\ln{${\rm L_{HX,nuc}}\,\,$}
\def\ba{\hskip-.1truecm}
\def\lb{${\rm L_{B}}\,\,$}
\def\lbs{${\rm L_{B\odot}}\,\,$}
\def\ledd{${\rm L_{Edd}}\,\,$}

\catcode`\@=11
\def\gsim{\ifmmode{\mathrel{\mathpalette\@versim>}}
    \else{$\mathrel{\mathpalette\@versim>}$}\fi}
\def\lsim{\ifmmode{\mathrel{\mathpalette\@versim<}}
    \else{$\mathrel{\mathpalette\@versim<}$}\fi}
\def\@versim#1#2{\lower 2.9truept \vbox{\baselineskip 0pt \lineskip
    0.5truept \ialign{$\m@th#1\hfil##\hfil$\crcr#2\crcr\sim\crcr}}}
\catcode`\@=12

\begin{document}

   \title{The X-ray emission properties and the
dicothomy in the central stellar cusp shapes of early-type galaxies}

   \author[S. Pellegrini]
          {S. Pellegrini\thanks{E-mail:silvia.pellegrini@unibo.it}
           \\
		   Astronomy Department, University of Bologna, 
                       via Ranzani 1, 40127 Bologna, Italy
          }
\date{Received ...; accepted ...}

   \maketitle

\begin{abstract} 

{\it Hubble Space Telescope} revealed a dicothomy in the central
surface brightness profiles of early-type galaxies, that subsequently
have been grouped into two families: core, boxy, anisotropic systems
and cuspy ("power law"), disky, rotating ones.  Here we investigate
whether a dicothomy is present also in the X-ray properties of the two
families. We consider both their total soft emission (\lt\ba), that is
a measure of the galactic hot gas content, and their nuclear hard
emission (\ln\ba), mostly coming from $Chandra$ observations, that is
a measure of the nuclear activity. At any optical luminosity, the
highest \lt values are reached by core galaxies; this is explained
with their being central dominant galaxies of groups, subclusters or
clusters, in many of the log \lt (erg s$^{-1})\gsim 41.5$ cases. The
highest \ln values, similar to those of classical AGNs, in this sample
are hosted only by core or intermediate galaxies; at low
luminosity AGN levels, \ln is independent of the central stellar profile
shape.
The presence of optical nuclei (also found by $HST$) is unrelated with
the level of \ln\ba, even though the highest \ln are all associated
with optical nuclei. The implications of these findings for the galaxy
evolution and the accretion modalities at the present epoch are
discussed.

\end{abstract}


\section{Introduction} 

{\it Hubble Space Telescope} observations of the central surface
brightness profiles $I(R)$ of nearby early-type galaxies revealed that the
profile shape has a bimodal distribution: either $I(R)$
breaks internally to shallower shapes $I\propto R^{-\gamma}$, with
$\gamma < 0.3$, and the corresponding galaxies are said to have cuspy
cores; or $I(R)$ follows a steep featureless power law that lacks a
core down to the resolution limit, with a slope $\gamma > 0.5$
(Ferrarese et al. 1994, Lauer et al. 1995, Faber et al. 1997; see also
Trujillo et al. 2004).  The core profile is typical of the most
luminous (\lb$\gsim 2.5\, 10^{10}$\lbs\ba) galaxies, while the
featureless power law profile is typical of the least luminous ones;
at intermediate luminosities, core and power law galaxies coexist.
More recently, Ravindranath et al. (2001) and Rest et al. (2001) have
identified a few systems with intermediate cusp slopes ($0.3<
\gamma<0.5$) that are however relatively uncommon, and the overall
bimodal distribution of the central profile shapes appears to be
robust (Lauer et al. 2005).

The division into two distinct classes of central structure was
emphasized further when the dynamical and morphological properties of
the galaxies were also considered: core galaxies are slow rotators and
have boxy or elliptical isophotes, while power law galaxies are rapid
rotators with disky isophotes (Kormendy \& Bender 1996, Faber et
al. 1997). These differences hold even when comparing power law and
core galaxies of the same luminosity. All this suggested that the
origin of the inner cusps is closely linked to the formation and
subsequent evolution of the galaxies; for example, the merging of
galaxies of comparable mass was conjectured to be responsible for
creating cores and boxy isophotes, while the steep density cusps were
linked to dissipation and therefore, possibly, to gas rich mergers
or minor mergers with less massive companions (Faber et al. 1997; Naab
\& Burkert 2003).

Another major $HST$ result is the widely accepted notion that massive
black holes (MBHs) are ubiquitous in the centers of spheroids (e.g.,
Richstone et al. 1998) and that relationships exist between the MBH
masses and the luminosity, central stellar velocity dispersion and
central light concentration of their hosts (e.g., Tremaine et
al. 2002, Graham et al. 2001). Therefore, there seems to be also a
deep relationship between MBHs and their associated galaxies. In this
context, many efforts have recently been made to link the creation of
cusps and cores to the effects of MBHs, as they should have
substantial influence on the dynamics and evolution of the surrounding
gas and stars (e.g., Cipollina \& Bertin 1994, van der Marel 1999,
Milosavljevic et al. 2002).  A popular scenario currently figures out
that during the merging of galaxies, each harboring a central MBH,
dynamical friction causes the MBHs to form a binary; interactions
between the binary and the surrounding stars (or gas) would harden the
binary until its coalescence, with ejection of stars from the center
and the production of a core.  

In this paper we investigate whether a dicothomy is present also in
the X-ray properties of core and power law galaxies. In fact both the
origin of the inner cusps, closely linked to the evolution of
galaxies, and the presence of a MBH are expected to have an
influence on the X-ray properties. In particular, the
information coming from this wavelenght allows us to investigate
the relationship of $\gamma$ with the galactic hot gas content 
(that manifests itself in the soft X-ray band; e.g., Kim et al. 1992) and
with the nuclear AGN-like properties, that dominate in the hard
X-ray band (Loewenstein et al. 2001, Pellegrini 2005).  In an earlier
work, Pellegrini (1999) investigated whether power law and core
galaxies differ systematically in their soft X-ray luminosities.  By
using the data available at that time, coming from WFPC1, it was found
that core galaxies can have a largely varying hot gas content, from
being devoid of hot gas to being gas rich; on the contrary, power law
galaxies are all confined to a more modest hot gas content. This held
even in the range of optical luminosities where the two families
coexist, and therefore the soft X-ray emission was considered
another property distinguishing the two families of core and power
law galaxies. Since then, the inner stellar light profiles have been
observed again with the improved resolution of WFPC2, and larger
samples with more homogeneous measurements of $\gamma$ have been
produced (Rest et al. 2001, Ravindranath et al. 2001, Lauer et
al. 2005).  As a result of the improved central profiles'
characterization, a number of galaxies have been reclassified.  

Here the problem of the relationship between the global hot gas
content and the inner stellar cusps is revisited by taking advantage
of the new samples of galaxies with $\gamma$ measured, as well as of
larger samples with the soft X-ray emission measured (O'Sullivan et
al. 2001a).  In addition, we can now investigate for the first time
whether there is a relation between the central structure of galaxies
and the presence or absence of nuclear activity. In fact, accurate
measurements of the nuclear X-ray emission have become available
recently thanks to $Chandra$ observations (e.g., Pellegrini 2005) and
a deep exploitation of the archival $ROSAT$ HRI observations (Roberts
\& Warwick 2000, Liu \& Bregman 2005).  Finally, in a substantial
fraction of the photometrically investigated galaxies $HST$ showed
nuclei, i.e., compact light sources that rise above the inward
extrapolated surface brightness cusp at small radii, as residuals of
the $I(R)\propto R^{-\gamma}$ law fits (Lauer et al. 1995,
Ravindranath et al. 2001, Rest et al. 2001). The origin of these
optical nuclei is still uncertain; their relationship with the nuclear
activity is investigated here by using the nuclear hard X-ray
emission.

\section{The sample}

All early-type galaxies with ``Nuker law'' parametric fits of their inner
brightness profiles observed with $HST$ were collected.  In this fit
the data are modeled with a double power law with a break radius, and
the value of the inner surface brightness slope (the $\gamma $
parameter) gives the classification into core or power law galaxy
(e.g., Faber et al. 1997). In a few cases the central profile can even
be declining ($\gamma<0$, Lauer et al. 2005); few galaxies have
$0.3<\gamma<0.5$ and are termed intermediate. There is no specific
criterion that characterizes this sample of galaxies with central
brightness parameters, in general it
comprises relatively luminous nearby ellipticals and S0s.  The
original work by Faber et al. (1997) made use of WFPC1 observations;
in more recent times, most of these galaxies were re-observed with
WFPC2.  Therefore, when a galaxy has been studied more than once, its 
$\gamma$ is taken here following this priority: first the sample
of 77 galaxies imaged with $HST+$WFPC2 of Lauer et al. (2005) is
considered; next that of 67 galaxies observed with $HST+$WFPC2 and
studied by Rest et al. (2001); next that of 61 galaxies by Faber et
al. (1997); next that of 33 galaxies observed with $HST+$NICMOS
by Ravindranath et al. (2001); finally, the $HST+$NICMOS sample by
Quillen et al. (2000) and the WFPC1-based works of Crane et al. (1993)
and Ferrarese et al. (1994).  

For this total sample of galaxies with information on the inner shape
of the optical profile, the literature has been searched for the
global soft X-ray emission (hereafter \lt\ba) and the nuclear hard
emission in the 2--10 keV band (hereafter \ln\hskip-.1truecm). Almost
all galaxies have \lt listed in the large catalogue based on $ROSAT$
PSPC observations, that were sensitive over 0.1--2.4 keV, of
O'Sullivan et al. (2001a).  This catalogue includes all the early type
galaxies with a Virgo corrected recession velocity $v\leq 9000$ km
s$^{-1}$ and apparent magnitude $B_T\leq 13.5$, and gives their \lt
derived by fitting the data with a MEKAL hot plasma model of
temperature $kT=1$ keV and solar metal abundance. For 8 galaxies \lt
comes from other studies (based on $ROSAT$ data in 6
cases\footnote{These are: NGC524 (Heldson et al. 2001), NGC4494
(O'Sullivan \& Ponman 2004), NGC4594 (Fabbiano \& Juda 1997), NGC4874
and NGC4881 (Dow \& White 1995), IC4329 (Irwin \& Sarazin 1998).} and
on $Einstein$ data for NGC2841 and NGC4342, Fabbiano et al. 1992).

\begin{figure*} 
\vskip -11truecm
\hskip -0.89truecm
\includegraphics[height=0.9\textheight,width=1.05\textwidth]{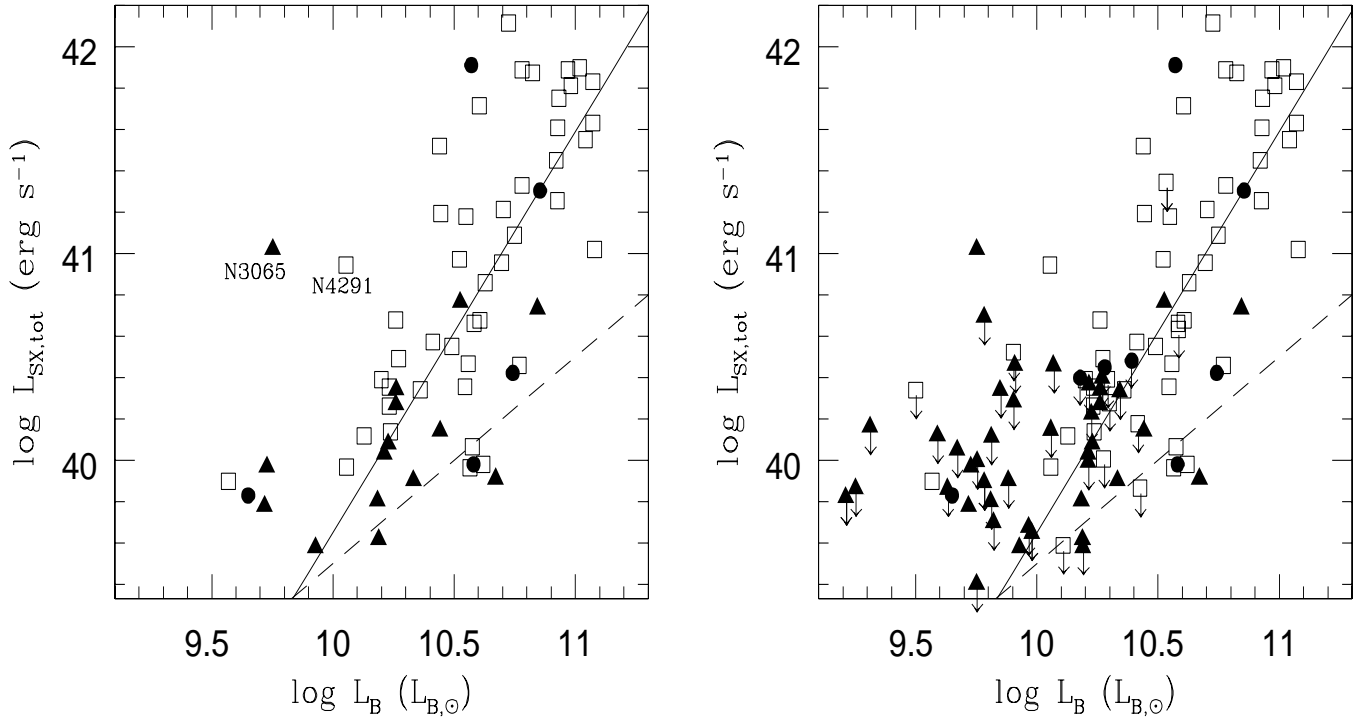}
\caption{The total soft X-ray luminosity \lt versus the total B-band
luminosity \lb for early-type galaxies with inner surface brightness
profile measured by $HST$ (detections only in the left panel,
all data in the right panel, with upper limits on \lt shown with a downward
arrow).  Full triangles indicate power law galaxies, full circles
intermediate ones and  open squares core ones. The solid line is the best 
fit \lt--\lb
relation for early-type galaxies and the dashed line is an estimate
for the contribution to \lt from stellar sources (O'Sullivan et
al. 2001a). Table 1 includes also the core galaxies NGC404 and 5 others
with log \lt (erg s$^{-1})>42.2$ (all at log \lb(\lbs\ba)$\geq
10.86$), and 5 power law galaxies with log \lt (erg s$^{-1})<39.4$ (all
at log \lb(\lbs\ba)$<9.7$); these do not appear in the plot due to the
boundary limits chosen to better show the region where both galaxy
types coexist.  %\label{fig1} 
} 
\end{figure*}

The nuclear hard luminosity \ln derives instead from a variety of
works.  Care was taken to draw as many measurements as possible
from $Chandra$ pointings, that have the best angular resolution 
so far ($\sim
0^{\prime\prime}\hskip-0.1truecm .5$); this was the case for most of
the nuclei (34). Next, $ROSAT$ HRI pointings have been considered
(with an angular resolution of $\sim 5^{\prime\prime}$) and these gave
\ln for 17 galaxies; finally, $XMM-$Newton pointings gave \ln for
NGC5548 and NGC7213, and $ASCA$, with a much larger angular resolution
($\sim 3^{\prime}$), gave \ln for 3 cases  (NGC3065, NGC3607, NGC7743),
where the nuclei are
considered to dominate the emission. All
nuclear luminosities have been converted to the 2--10 keV band by
using the spectral parameters found by the authors who analyzed the
X-ray data.

Table 1 lists the 116 galaxies with both $\gamma$ and \lt measured.
This sample is almost double as large as that considered by
Pellegrini (1999), that included 59 objects.  In addition, the present
sample differs from the old one also because of some new features: it
includes a few core galaxies at log \lb(\lbs\ba)$<10.2$ and the
intermediate galaxies, a few power law galaxies of the old sample have
been re-classified as core or intermediate galaxies, and a few X-ray
upper limits in the old sample are now detections. Table 2 lists the 56
galaxies for which a search for the nuclear emission has been performed
(5 of these are not in Table 1).

The Tables also list the blue luminosities \lb of the galaxies,
taken from O'Sullivan et al. (2001a) or calculated from the $B_T$
magnitudes of the LEDA catalogue, as done by O'Sullivan et
al. (2001a). The distances used here are instead those
homogeneously derived from the SBF method of Tonry
et al. (2001). When this kind of distance is not available, the Lauer
et al.'s or the O'Sullivan et al.'s distances or the distance used for
the X-ray analysis (in this order) are adopted, after reascaling for
the $H_0=74$ km s$^{-1}$ Mpc$^{-1}$ (as implied by
Tonry et al. 2001). The \lt\hskip-.1truecm, \ln and
\lb values in Table 1 and 2 have been rescaled for the distances
adopted here.

\section{Results}

\subsection{\lt and inner light profile}

In Fig. 1 the \lt-- \lb relation for the galaxies in Table 1 is plotted,
with different symbols for core, intermediate and power law systems.
As found in general (Sect. 1), core galaxies are more frequent at
higher \lb\ba, while power law ones are more common at lower \lb\ba.

Figure 1 shows that the most X-ray luminous galaxies are core and
intermediate ones; also, for log \lb(\lbs\ba)$> 10.4$ these cover the
whole range of observed \lt values, that extends for roughly two
orders of magnitude.  Power law systems are instead confined below log
\lt (erg s$^{-1})\sim 41$; they tend to be less luminous than core
ones at every L$_{\rm B}$\hskip-.01truecm,
with a marked difference at log \lb(\lbs\ba)$>
10.4$.

A statistical analysis has been performed to quantify how strong is
this apparent dicothomy in the soft X-ray properties of core and power
law galaxies. A series of Kolmogorov-Smirnov and Two Sample Tests
(contained in the ASURV package, Feigelson \& Nelson 1985) have been
applied to establish first whether core and power law galaxies are
consistent with being drawn from the same \lb distribution, and then
whether they are also consistent with the same \lt distribution.  Two
intervals of \lb values have been considered: a larger one where the
two families overlap [10.0 $<$ log \lb(\lbs\ba) $<$ 10.7], and the
10.0 $<$ log \lb(\lbs\ba) $\leq 10.4$ region, where there is an equal
number of core and power law galaxies (15 each).  In the large \lb
interval, the hypothesis of the same \lb distribution can be excluded
at the $\sim 2.4\sigma$ level only, while that of the same \lt
distribution is excluded at the $\sim 3.2\sigma$ level. In the small
\lb interval, the two families are definitely consistent with having
the same \lb distribution, while the hypothesis of the same \lt
distribution can be excluded at the $\sim 2.6\sigma$ level.

At log \lb(\lbs\ba)$>10.4$, where the dicothomy in \lt is more
evident, there are in Fig. 1 many more core systems than power law
ones, and therefore the confinement of the latter to lower \lt could
be coincidental. However, all the power law galaxies still lacking a
measurement of \lt have log \lb(\lbs\ba)$\leq 10.2$. Consequently,
Fig. 1 is already representative of the soft X-ray properties of the
power law family presently known.  Note also how there are just 4
intermediate galaxies at log \lb(\lbs\ba)$>10.4$, yet these show a
range of \lt fully comparable to that of core systems.  Therefore,
intermediate galaxies share the same soft X-ray properties of core
ones.

Finally, note that the highest \lt value of
power law galaxies (that of NGC3065, evidenced in Fig. 1) can be mostly
ascribed to nuclear activity\footnote{The predicted {\it nuclear}
luminosity over 0.1--2.4 keV, obtained by using the Iyomoto et
al. (1998) spectral shape (Table 2), is $\sim $\lt derived by
O'Sullivan et al. (2001a). On the contrary, the core galaxy NGC4291
close to NGC3065 (Fig. 1) has \lt mostly due to a hot ISM, since its
nuclear emission accounts for $\sim 0.15$ of \lt (for the spectral
shape of the reference in Table 2).}.  Therefore the soft X-ray
emission due to hot gas is really confined below $\sim 10^{41}$ erg
s$^{-1}$ for all power law galaxies.  On the other hand, the
\lt$>10^{41}$ erg s$^{-1}$ values of most core and intermediate
galaxies cannot be attributed totally or substantially to nuclear
activity, given the generally low or very low emission level of the latter
(Sect. 3.2, Table 2).

\subsection{\ln and inner light profile}

\begin{figure}
\vskip -12truecm
\hskip -0.5truecm
\includegraphics[height=.9\textheight,width=1.05\textwidth]{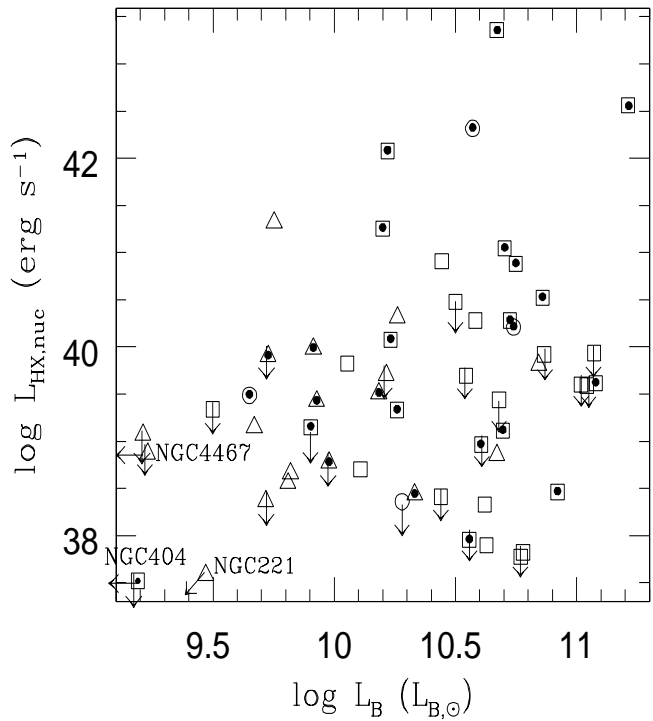}
\caption{The 2--10 keV nuclear luminosity \ln versus the galactic \lb
for early-type galaxies with inner surface brightness
profile measured by $HST$.  Power law galaxies are shown with 
triangles, intermediate ones with circles and core ones with
 squares. Upper limits on \ln are shown with a
downward arrow. The data used are those in Table 2. Galaxies with optical
nuclei have a dot inside their symbol. NGC221 would be located below and to the
left of the plot; NGC404 and NGC4467 to the left (their \ln are upper limits). 
\label{fig2}}
\end{figure}

\begin{figure*} 
\vskip -11truecm
\hskip -0.89truecm
\includegraphics[height=0.9\textheight,width=1.05\textwidth]{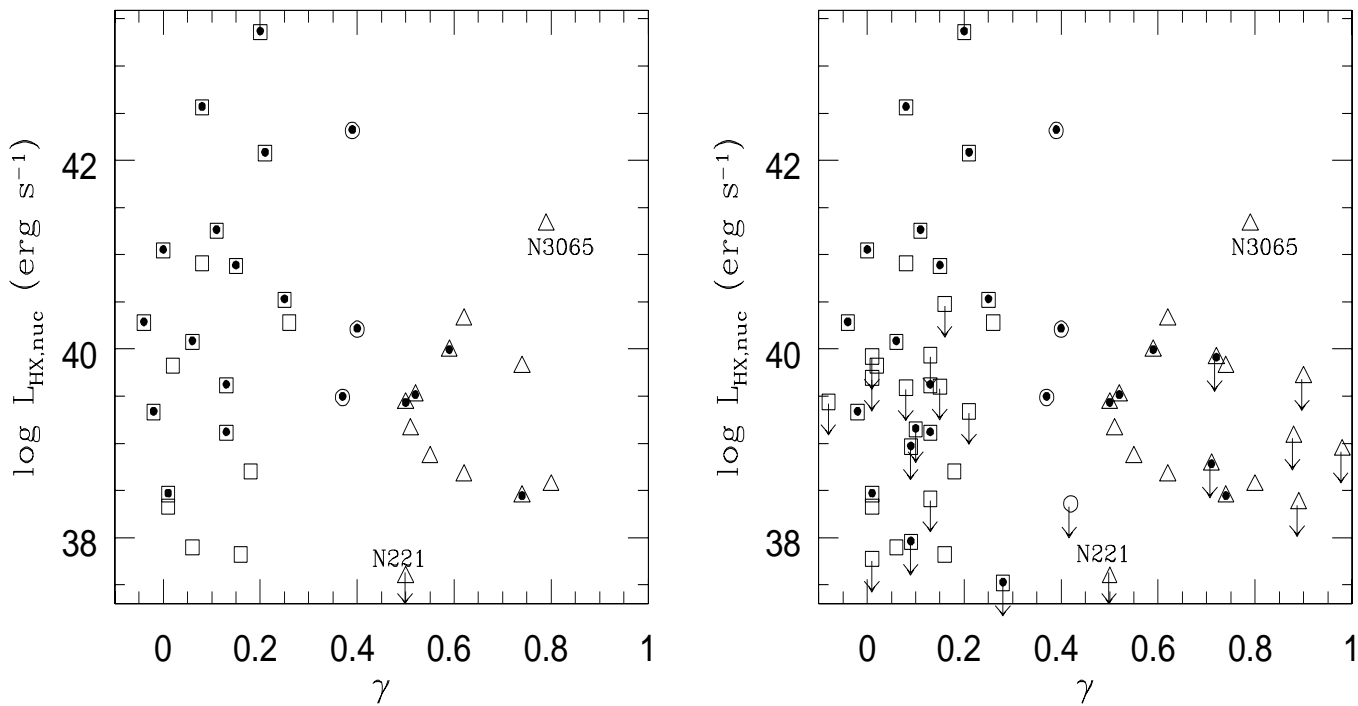}
\caption{The hard nuclear X-ray luminosity \ln versus the shape
parameter $\gamma$ for the galaxies in Table 2 (detections only in 
the left panel, all data in the right one, with upper limits on \ln
shown with a downward arrow). Power law
galaxies are shown with triangles, intermediate ones with
circles and core ones with squares. Galaxies with an
optical nucleus  have a dot inside their symbol. The location of 
NGC221 is below the plot (it is not an upper limit).
The four objects with the highest \ln are
NGC3862, NGC5548, NGC6166 and NGC7213.
\label{fig3}}
\end{figure*}

Figure 2 shows the relationship between the hard nuclear emission \ln
(from Table 2) and the galactic \lb\ba, for
core, intermediate and power law galaxies. This figure indicates that
the galaxies with known \ln cover the same large range of \lb values of
Fig. 1.
The relationship between \ln and the inner optical
profile shape $\gamma$ is plotted in Fig. 3, which shows that:

1) core and intermediate galaxies span the whole range of \ln values
observed, from those typical of the faintest detected nuclei up to
values typical of ``classical'' AGNs as Seyfert galaxies. 

2) power law systems only host low luminosity AGNs, i.e., nuclei 
with \ln$\lsim $ few$\times  10^{41}$ erg s$^{-1}$ (e.g., Terashima et al. 
2002). Their highest \ln is that of NGC3065 (evidenced in  Fig. 3); since it 
derives from ASCA data, it could have been somewhat overestimated, 
but a low luminosity AGN is certainly present in this galaxy, given that
it hosts a LINER with broad Balmer lines (Eracleous \& 
Halpern 2001) and its \ln is $10-100$ times higher than expected from the X-ray
binaries falling in the extraction region used for the spectrum (Iyomoto et
al. 1998).

3) in the realm of low luminosity AGNs, 
core and power law profiles cover the same range of \ln values, and
there is no clear trend between $\gamma$
and \ln\hskip-.1truecm.

4) the Eddington ratio \ln\ba/\ledd is also independent of $\gamma$
(Fig. 4). Here \ledd is derived from the $M_{BH}-\sigma$ relation
(e.g., Tremaine et al. 2002), where $\sigma $ is the central stellar
velocity dispersion (from McElroy 1995). \ln\ba/\ledd is close to unity
only for the Seyfert NGC5548, while it is $\lsim 10^{-4}$ for all the other
nuclei.

We note that power law galaxies seem to host nuclear emission
less frequently than core ones: in Table 2 there are 34 core systems
(20 of which are detections) versus 18 power law ones (12 detections).
This core/power law proportion is more unbalanced than in Table 1
(where it is 61/47).  However, at present we cannot decide whether
power law galaxies are less frequently active or they have been chosen
less frequently as targets. One reason of preference for pointing core
galaxies could be that they are on average brighter; another could be
that they include interesting targets as centers of groups/clusters,
that in turn generally host activity at a detectable level (see also
Sect. 4).

\begin{figure*} 
\vskip -11truecm
\hskip -0.89truecm
\includegraphics[height=0.9\textheight,width=1.05\textwidth]{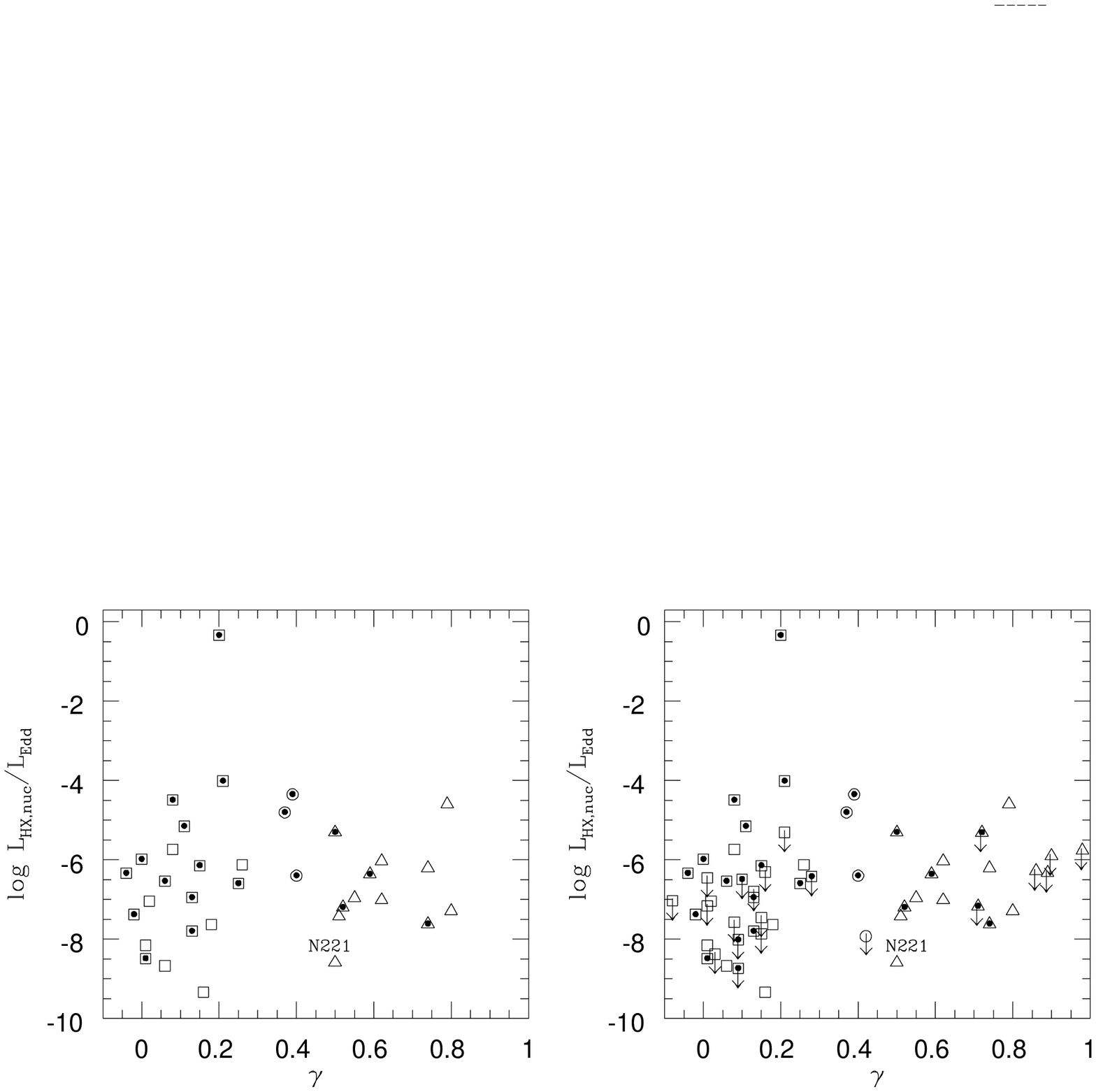}
\caption{The trend of \ln scaled by the Eddington
luminosity versus the $\gamma$ parameter (see Sect. 3.2). Symbols and data
are the same as in the previous Fig. 3, with detections only in the left 
panel, all
data in the right one. The only one object with an Eddington ratio close
to unity is NGC5548.
\label{fig4}}
\end{figure*}

\subsection{\ln and presence of optical nuclei}

As mentioned in the Introduction, many galaxies show nuclei, 
compact light sources that rise above the inward extrapolated surface
brightness cusp at small radii.  In general these nuclei are bluer
than the background starlight and spatially unresolved. They could be
nuclear star clusters, in which case they may comprise stars younger
or more metal poor than those surrounding the nuclei, or they could be
low luminosity AGNs. Ravindranath et al. (2001) argued that the majority
of their nuclei are associated with AGNs; Lauer et al. (2005)
found nuclei in 29\% of core galaxies and 60\% of power law ones, with
weak evidence that nuclei in power law galaxies have absorption line
spectra while those in core galaxies have emission lines. However,
since they also found core galaxies with emission lines and no nuclei,
the nature of these nuclei remained unkown.  

In the sample considered in Table 2 the presence of optical nuclei has
been derived from the references giving the slope $\gamma $.  They are
found in $\sim $half of the core and $\sim 40$\% of the power law
galaxies.  As Figs. 2, 3 and 4 show, optical nuclei are present at all
\lb values and, most importantly, at all levels of X-ray activity.
Also, there is not a strong relationship between the level of \ln and
the frequency of nuclei, except for the fact that all the highest \ln
are associated with optical nuclei.  It seems reasonable then to
conclude that some nuclei are likely associated with nuclear activity,
for the others alternative origins are equally probable.

\section{Discussion}

\subsection{A dicothomy in hot gas content}

The main finding concerning the \lt properties of core and power law
galaxies is that at any \lb core systems cover the whole observed
range of \lt values, while power law systems tend to be underluminous
in \lt with respect to core ones, especially at high \lb\ba.  If we
read \lt as a measure of the hot gas content (e.g., Kim et al. 1992),
then core galaxies are on average richer of hot gas at any 
L$_{\rm B}$\hskip-.04truecm, and
can be very much richer at high \lb\ba. We discuss here whether these
findings are directly linked to the slope of the stellar profile in the
inner galactic region or are the result of other different properties
of the two families.

Hydrodynamical simulations of the hot gas evolution in early type
galaxies give an estimate of their ``normal'' hot gas content, i.e.,
that accumulated from stellar mass losses in an isolated galaxy during
its lifetime.  A set of simulations for spherical galaxy models with
central stellar density distributions of the shape found by $HST$ was
run by Pellegrini \& Ciotti (1998). A reason for a different hot gas
content of core and power law galaxies was not found, since the {\it
galactic centers} are almost always dense enough to host gas inflows,
regardless of differences in the density profile shapes at the $HST$
resolution limit. Instead, another one of the fundamental properties
defining the core and power law families likely affects the gas
content on the {\it galactic scale}: power law systems are typically
more flattened than core ones of the same \lb (Sect. 1; Kormendy \&
Bender 1996, Lauer et al. 2005), and flatter systems are
systematically underluminous in X-rays with respect to rounder ones,
at fixed \lb (Eskridge et al. 1995). This was explained
by a flatter mass distribution
corresponding to a shallower potential well, which makes it easier for
 the hot gas to escape from the galaxy (Ciotti \& Pellegrini 1996,
Pellegrini et al. 1997, D'Ercole \& Ciotti 1998).

This "global shape" effect can account for the $\sim 2.6\sigma$
difference in \lt shown by the two families at lower \lb (Sect. 3.1),
but it cannot fully explain the very large difference in \lt at high
\lb\ba. Actually, the highest \lt values of core galaxies cannot be
justified within the framework of the above mentioned models.  In fact
the maximum \lt at each \lb increases with \lb\ba, since the hot gas
is more bound in galaxies that have on average deeper potential wells
(e.g., Ciotti et al. 1991), but many core galaxies exceed the maximum
values predicted by the models. An additional contribution to their
\lt may then come from a dense intragroup medium (IGM) or intracluster
medium (ICM), not fully subtracted during the data analysis. This
subtraction is particularly problematic for galaxies at the center of
groups and clusters (O'Sullivan et al. 2001a), and in fact the core
galaxies with the highest \lt are in most cases central members of
large groups, subclusters or clusters (NGC507, NGC741, NGC1399,
NGC3842, IC4329, NGC4073, NGC4486, NGC4889, NGC5419, NGC6166, NGC7619,
NGC7768).  Instead, no power law systems residing at the center of gas
rich groups or clusters are present in the sample considered here.
This is in line with the findings for a large sample of central
cluster galaxies: their inner light profiles turned out to be typical
of core galaxies, with power law profiles only in 10\% of them (Laine
et al. 2003)\footnote{No \lt or \ln are available in the literature
for these six power law galaxies.}.  In addition, central dominant
galaxies have different X-ray properties compared to non-central
members or field galaxies (Helsdon et al. 2001, Matsushita 2001): they
show an overluminosity and an extension explained by their hot gas
content being more closely correlated
with the properties of the group/cluster as a whole than with those of the
single galaxy, and also by their being often at the center of
a group/cluster cooling flow (e.g., Fabian 1994).

Another significant contribution to \lt may come from nuclear
activity, as possible for those four
core/intermediate objects with the highest \ln of Fig. 3 (see also the
next Section). In conclusion, the very large difference in \lt at high
\lb is due to properties that belong preferentially to core galaxies,
as being central dominant galaxies and/or hosting a bright AGN.

\begin{figure} 
\vskip -1truecm
\hskip -1.truecm
\includegraphics[height=0.55\textheight,width=0.7\textwidth]{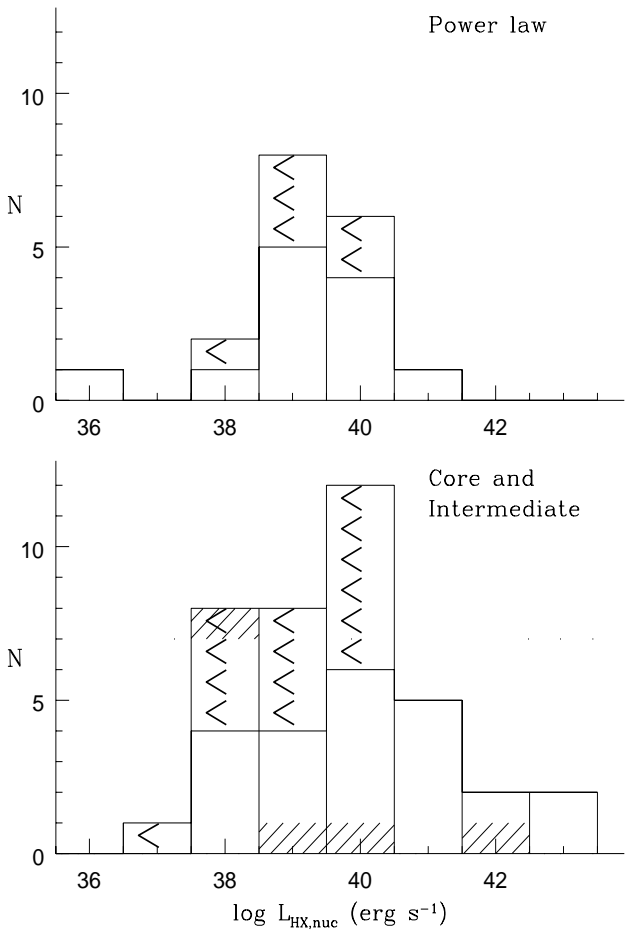}
\caption{The distribution of \ln 
for power law galaxies (upper panel) and
for core plus intermediate galaxies (the latter are four and
are indicated with shading; lower panel).
Upper limits on \ln are indicated with a "$<$".
}
\end{figure}

\begin{figure} 
\vskip -1truecm
\hskip -1.truecm
\includegraphics[height=0.55\textheight,width=0.7\textwidth]{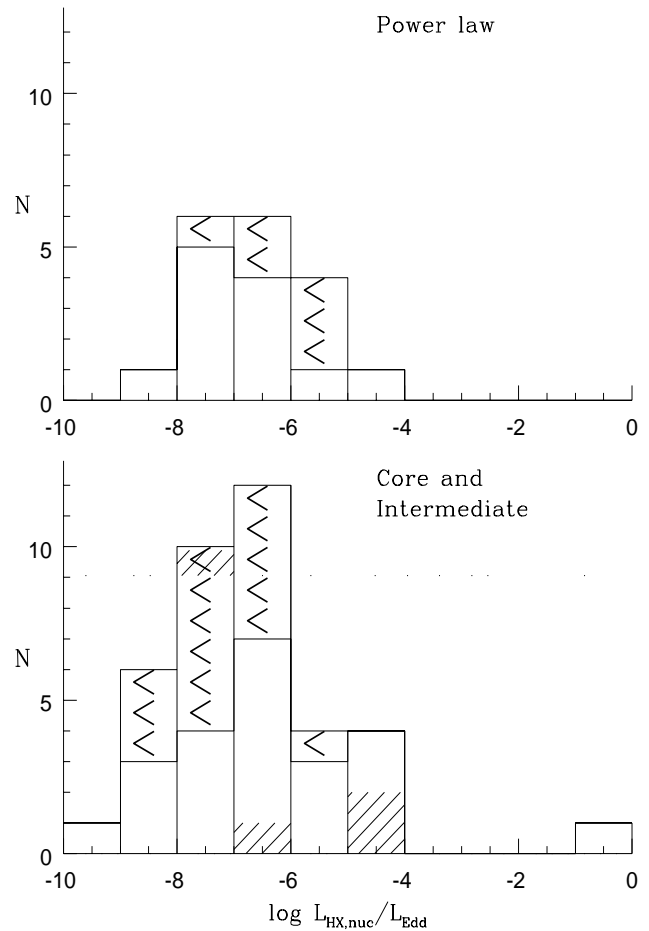}
\caption{The distribution of \ln\ba/\ledd with symbols as in the
previous Fig. 5. }
\end{figure}

\subsection{A dicothomy in nuclear activity?}

The dicothomy found here is that the brightest \ln values, at the
level of classical AGNs, are found only for core or intermediate
galaxies. This conclusion could actually be hampered by small number
statistics; however, a trend of this kind is somewhat expected:
nuclear activity is favoured by being the host galaxy the bright
central member of a group or cluster (e.g., Burns 1990; here, among
the brightest \ln galaxies, NGC6166 resides at the center of Abell
2199 and NGC3862 lies in a dense part of Abell 1367), and the vast
majority of central dominant galaxies are core systems (Laine et
al. 2003).  Actually, it can be concluded that high \ln values are not
found in power law galaxies for a sample larger than that in Table 2,
as long as nuclear activity is not heavily absorbed and gives a large
contribution also to \lt\ba, as in Type 1 AGNs (e.g., Antonucci 1993).
For example, in NGC3065 most of \lt comes from a low luminosity AGN
(Sect. 3.1); but this remains the highest \lt case of power law
galaxies.

Lower activity levels [log \ln (erg s$^{-1})\lsim 41.3$] are instead
equally common for all $\gamma $ values. This is in line with
the current belief that all spheroids host a central MBH (Richstone et
al. 1998), independently of the shape of their inner light profile.
What was not expected a priori is that this level of nuclear emission
is unrelated with the optical profile shape, that consequently
does not play any direct role in the feeding of the central MBH, 
either in the sense of favouring or opposing it.

Other aspects concerning a possible dicothomy deserve to be explored
with a larger sample of power law galaxies. For example, it is to be
established whether they really never reach the highest \ln values of
core systems, and whether they have a distribution of \ln and
\ln\ba/\ledd significantly different from that of core galaxies (see
Figs. 5 and 6). For the data available, a series of Two Sample Tests
as those mentioned in Sect. 3.1 shows that the \ln and \ln\ba/\ledd
values of the two families are consistent with being
drawn from the same distribution.  Another aspect to be investigated
is whether power law galaxies are X-ray active less frequently than
core ones, as mentioned in Sect. 3.2. Such a tendency is revealed also
when nuclear activity is measured by the optical line emission (of the
LINER, Seyfert or Transition type; e.g., from Ho et al. 1997).
Ravindranath et al. (2001) reported a marginally higher detection
frequency of nuclear line emission among core type systems, and
optical activity is found in $\sim $half of the core galaxies and
$\sim 1/3$ of the power law ones in the Lauer et al. (2005) sample
(from their Tab. 3).  In the sample considered here (Tables 1 and 2),
$\sim 60$\% of core and $\sim 1/3$ of power law galaxies show optical
activity; therefore, in the optical core
systems tend to be active more frequently than power law ones.

\subsection{Relationship with the galaxy evolution}

A few final considerations are presented here to relate the supposed
origin and evolution of the two families of core and power law
galaxies (described in Sect. 1) with their X-ray properties discussed
in this work.

Fig. 3 reminds of the behavior of boxy and disky galaxies with respect
to the radio luminosity: boxy galaxies span a large range of radio
power values, while disky galaxies are found only at low radio powers
(Bender et al. 1989). Given that the radio luminosity and \ln both
trace the nuclear activity, this similarity is expected, because boxy
galaxies are more frequent among cores systems and power law galaxies
among disky ones (Sect. 1).  Since the work of Bender et al., the link
between boxiness and activity has been explained by 
boxiness being generally associated with anisotropic (triaxial) systems,
where it is easier for the ISM to reach the nucleus than in more
rapidly rotating ones.  Another suggestion was that boxy and irregular
structures are a result of merging processes or various types of
interactions (Nieto \& Bender 1989), that in turn also seem to trigger
nuclear activity (see, e.g., Martini 2004 for a review).

It is unlikely though that the same merging episod created both the
fundamental properties of the family of core galaxies (the low
rotational level, the boxy isophotes and the ejection of stars from
the center with the production of a core) and the onset of the nuclear
activity that is still observed today (Fig. 3). In fact the period of
activity is $\lsim 10^8$ yrs (e.g., Martini 2004), a time that is
uncomfortably close to that required for the merging product to reach
equilibrium ($\sim $few dynamical times, that is $\sim$few $10^8$ yrs
in the central galactic regions).  Also, nuclear activity {\it of low
level} seems unrelated with the fundamental properties defining the
two families of core and power law galaxies (i.e., $\gamma$ from
Figs. 3 and 4; global isophotal shape from the study on the radio
emission by Bender et al.  1989) and likewise is unrelated with the
MBH mass and the mass accretion rate on it, estimated under steady
state hypotheses (Pellegrini 2005).  A comprehensive explanation could
be that activity follows cycles of on and off periods, not influenced
by long lasting and global properties of the galaxies (Binney \& Tabor
1995, Ciotti \& Ostriker 2001, Omma et al. 2004, Sazonov et
al. 2005). In this context, few systems are expected to show a large
\ln value, at the present epoch.

Another aspect concerns the way in which merging affects the
galactic hot gas content.  In an observational study of the X-ray
evolution of on-going mergers, the late stages were found to be
underluminous in \lt compared with the typical values for early-type
galaxies of the same \lb (Read \& Ponman 1998; see also Fabbiano \&
Schweizer 1995, O'Sullivan et al. 2001b).  Although when two galaxies
coalesce massive hot extended gas is observed, after this time \lt
decreases, and the late, relaxed remnants appear devoid of gas.  If
major mergers are the progenitors of normal ellipticals, the X-ray
halo of hot gas must be regenerated. The most plausible mechanism for
such a replenishment is through mass losses from evolving stars
(Ciotti et al. 1991, O'Sullivan et al. 2001b); this requires many
Gyrs to produce the massive ($\sim 10^9-10^{10}M_{\odot}$) hot halos
of the core galaxies with the highest \lt in Fig. 1. Therefore, these
must have had their last major merger many Gyrs ago.

\section{Conclusions}

Core and power law galaxies show a clear dicothomy of properties in
the soft X-ray emission, in the sense that core galaxies tend to be
more hot gas rich. The main dicothomy in the nuclear hard X-ray
emission is the lack of bright nuclei among power law galaxies (to be
tested with a larger sample of the latter).  More in detail, the
results of this work can be summarized as follows:

\begin{enumerate} \renewcommand{\theenumi}{\arabic{enumi}.}

\item At any \lb core galaxies cover the whole range of observed \lt
values; this can be very large ($\sim 2 $ orders of magnitude) at
log \lb$(L_{\odot})> 10.4$, reaching \lt$\gsim 10^{42}$ erg s$^{-1}$.
Power law galaxies tend to be underluminous in \lt with
respect to core ones at every \lb\ba, and especially at high \lb\ba.

\item The above properties are not directly resulting from the shape
of the inner optical profile. The underluminosity of power law
galaxies has a contribution from their average flatter mass
distribution, which favors the gas outflow. The large
overluminosity in \lt of core systems of high \lb is often linked to
their being central dominant galaxies. Instead, power law systems at
the center of hot gas rich groups or clusters are lacking from this
sample (as are not found in such a position in general).

\item The highest nuclear luminosities in the 2--10 keV band are
reached by core or intermediate galaxies.  In the low luminosity AGN
domain, \ln is independent of $\gamma $: core and power law profiles
correspond to the same large range of \ln\ba, with no trend with $\gamma $.
The Eddington ratio \ln\ba/\ledd is always very low ($\lsim 10^{-4}$),
except for one core galaxy, again without any trend with
$\gamma$.

\item Intermediate galaxies share the same \lt and \ln properties as core
ones.

\item The presence of nuclear hard emission seems more frequent
among core galaxies than power law ones, as  seems also to be
the case for optical line emission. 
 
 \item The presence of optical nuclei is unrelated with the level of
nuclear hard emission. The highest \ln\ba, though, are all associated
with optical nuclei.

 \item It is unlikely that the same merging episod was responsible for
 the building of the galactic structure (e.g., the isophotal shape,
 the $\gamma $ value) and the nuclear activity observed today. The
latter appears mostly of low level and unrelated with the global
 galaxy properties (including $\gamma$). Also the hot massive
 haloes of many core galaxies suggest that their last major merging
 episod took place many Gyrs ago.

\end{enumerate}

\begin{table}
\caption[] { Properties of galaxies with $\gamma$ and \lt measured}
\begin{flushleft}
\begin{tabular}{ l  r  r  r  r l  l }
\noalign{\smallskip}
\hline
\noalign{\smallskip}
 Name & d$^a$  & log \lb$^b$ & log L$_{\rm SX,tot}^c$ & $\gamma^{d} $ & Ref.\\
      & (Mpc)  & $(L_{B\odot})$    &(erg s$^{-1})$    &               &  \\
\noalign{\smallskip}
\hline
\noalign{\smallskip}
Core:\\ 
         NGC404           &        2.4           &      8.47           &       $<$38.28        &           0.28           &           Ra\\
         NGC507           &        60.3          &       10.87           &       42.61           &           0.00           &           L\\
         NGC524           &       24.0           &       10.43           &       $<$39.87           &        0.03           &           Ra\\
         NGC584           &       20.1           &       10.28           &       $<$40.01           &         0.30           &           L\\
         NGC720           &       27.7           &       10.63           &       40.86           &        0.06           &           F\\
         NGC741           &       67.1           &       10.98           &       41.81           &         0.10           &           L\\
         NGC1016           &       71.8           &       10.93           &       41.26           &        0.09           &           L\\
         NGC1052           &       19.4           &        10.20           &       40.39           &        0.11           &           Ra\\
         NGC1316           &       21.5           &       11.08           &       41.02           &        0.13           &           L\\
         NGC1374           &       19.8           &       10.06           &       39.97           &       -0.03           &           L\\
         NGC1399           &       20.0           &        10.60          &       41.71           &        0.09           &           L\\
         NGC1400           &       26.4           &       10.36           &       40.34           &           0.00           &           F\\
         NGC1600           &       60.8           &       11.04           &       41.55           &        0.08           &           F\\
         NGC1700           &       38.4           &       10.61          &       40.68           &        0.07           &           L\\
         NGC2300           &       28.8           &        10.44          &       41.19           &        0.08           &           L\\
         NGC2832           &       87.1           &       11.07           &       41.63           &        0.02           &           F\\
         NGC2841           &       13.0           &       10.62           &       39.98           &        0.01           &           F\\
         NGC2986           &       30.7           &       10.52           &       40.97           &        0.18           &           Re\\
         NGC3193           &       34.0           &       10.55           &       40.36           &        0.01           &           Re\\
         NGC3379           &       10.6           &       10.11           &       $<$39.59           &        0.18           &           L\\
         NGC3607           &        22.8           &       10.58           &       40.66           &        0.26           &           L\\
         NGC3608           &       22.9          &       10.24           &       40.14           &        0.17           &           L\\
        NGC3613           &       29.1           &       10.42           &       $<$40.18           &        0.04           &         Re\\
         NGC3640           &       27.0           &       10.57           &       40.06           &        0.03           &           L\\
         NGC3706           &        44.4           &       10.53           &       $<$41.34           &       -0.01           &           L\\
         NGC3842           &       91.8           &       11.02           &       41.90           &        0.15           &           L\\
         NGC4073           &       87.2           &        11.15          &       42.46           &       -0.08           &           L\\
         NGC4168           &       34.1          &       10.41           &       40.57           &        0.17           &           Re\\
         NGC4261           &       31.6           &        10.70          &       41.21           &           0.00           &           Ra\\
         NGC4278           &       16.1           &       10.23           &       40.35           &         0.10           &           L\\
         NGC4291           &       26.1           &       10.05           &       40.94           &        0.02           &           L\\
         NGC4365           &       20.4           &       10.56           &       40.47           &        0.09           &           L\\
         NGC4374           &       18.4           &        10.70          &       40.96           &        0.13           &           Ra\\
         NGC4382           &       18.4           &       10.77           &       40.46           &        0.01           &           L\\
        NGC4406           &       17.1           &       10.73           &       42.12           &       -0.04           &           L\\
        NGC4458           &       17.2           &       9.57         &        39.90          &        0.17           &           L\\
        NGC4472           &       16.3           &       10.92           &       41.45           &        0.01           &           L\\
        NGC4473           &        15.7           &       10.13           &       40.12           &        0.01           &           L\\
        NGC4476           &       17.2           &       9.50           &       $<$40.34           &        0.21           &           Fe\\
        NGC4478           &       18.1           &       9.90           &       $<$40.52           &         0.10           &           L\\
        NGC4486           &       16.1           &       10.86           &       42.96           &        0.25           &           F\\
        NGC4552           &       15.4           &       10.26           &       40.68           &       -0.02           &           L\\
        NGC4589           &       22.0           &       10.23           &       40.26           &        0.25           &           L\\
        NGC4636           &       14.7           &       10.44           &       41.52           &        0.13           &           Ra\\
        NGC4649           &       16.8           &       10.78           &       41.33           &        0.16           &           L\\
        NGC4709           &       35.3           &       10.49           &       40.55           &        0.28           &           L\\
        NGC4874           &       89.5           &       11.07           &       41.83           &        0.13           &           F\\
        NGC4889           &       89.5           &        11.20           &       42.77           &        0.05           &           F\\
        NGC5061           &       25.4           &       10.56           &       39.96           &        0.05           &           L\\
        NGC5077           &       30.6           &       10.27           &       40.49           &        0.23           &           Re\\
        NGC5198           &       35.5           &       10.29           &       $<$40.39           &        0.23           &           Re\\
        NGC5419           &          59.2           &       10.97           &       41.89           &        0.03           &           L\\
        NGC5576           &       25.6           &        10.30          &       $<$40.28           &        0.26           &           L\\
        NGC5903           &       33.9           &       10.58           &       $<$40.63           &        0.13           &           Re\\
        NGC5982           &       38.3           &        10.55        &       41.18           &        0.05           &           L\\
        NGC6166           &       110.1           &       11.21           &       43.94           &        0.08           &           F\\
        NGC6876           &       54.4           &       10.93           &       41.61           &           0.00          &           L\\
        NGC7619           &       53.0           &       10.82           &       41.87           &        0.01           &           L\\
        NGC7768           &       93.2           &       10.93           &       41.75           &           0.00        &           F\\
        IC1459           &       29.2           &       10.75           &       41.09           &        0.15           &           L\\
\noalign{\smallskip}
\hline
\end{tabular} 
\end{flushleft}
\end{table}

\begin{table}
\contcaption{}
\begin{tabular}{ l  r  r  r  r  l l }
\noalign{\smallskip}
\hline
\noalign{\smallskip}
 Name & d$^a$  & log \lb$^b$ & log L$_{\rm SX,tot}^c$ & $\gamma^{d} $ & Ref.\\
      & (Mpc)  & $(L_{B\odot})$    &(erg s$^{-1})$    &               &  \\
\noalign{\smallskip}
\hline
\noalign{\smallskip}
        IC4329           &        53.6           &       10.78          &       41.89           &        0.01           &           L\\
\noalign{\smallskip}
Intermediate:\\
         NGC821           &        24.1           &       10.28           &       $<$40.45           &        0.42           &           L\\
        NGC3585           &       20.0           &       10.58           &       39.98           &        0.31           &           L\\
        NGC3862           &        83.1           &       10.57           &       41.91           &        0.39           &           C\\
        NGC4594           &       9.8           &       10.74           &       40.42           &         0.40          &           C\\
        NGC5273           &       16.5           &        9.65           &       39.83           &        0.37           &           Ra\\
        NGC5831           &       27.2           &       10.18           &       $<$40.40           &        0.33           &           Re\\
        NGC5898           &       29.1           &       10.39           &       $<$40.48           &        0.41           &           Re\\
        NGC7626           &       53.0           &       10.85           &        41.30           &        0.36           &           Ra\\
\noalign{\smallskip}
Power law:\\
        NGC221 (M32)          &        0.81           &       8.46         &       37.87           &         0.50       &           Ra\\
        NGC596           &       21.7          &       10.19           &       $<$39.58           &        0.54           &           L\\
        NGC1172           &       21.5           &       9.85           &      $<$40.34           &        1.01           &           F\\
        NGC1351           &       21.0           &       9.91           &       $<$40.46           &        0.78           &           Q\\
        NGC1426           &        24.1           &       10.06           &       $<$40.15           &        0.56           &           L\\
        NGC1427           &       23.6           &       10.23           &       40.08           &        0.51           &           L\\
        NGC1439           &       26.6           &       10.22           &       $<$40.22           &        0.74           &           L\\
        NGC1553           &       18.5           &       10.84           &       40.73           &        0.74           &           Q\\
        NGC2434           &       21.6           &       10.26           &       40.27           &        0.75           &           L\\
        NGC2634           &       33.4           &       10.07           &       $<$40.46           &        0.81           &           Re\\
        NGC2685           &       16.1           &       9.81           &       $<$40.11           &        0.73           &           Ra\\
        NGC2778           &       22.9           &       9.59           &       $<$40.12           &        0.83           &           L\\
        NGC2974           &       21.5           &       10.26           &       40.34           &        0.62           &           L\\
        NGC3065           &       30.4           &       9.75           &       41.02           &        0.79           &           Re\\
        NGC3078           &       35.2           &       10.52           &       40.76           &        0.95           &           Re\\
        NGC3115           &       9.7           &       10.18           &        39.80           &        0.52           &           L\\
        NGC3377           &       11.2           &        9.82           &       $<$39.70           &        0.62           &           L\\
        NGC3384           &       11.6           &       9.98           &       $<$39.65           &        0.71           &           L\\
        NGC3599           &       20.3           &       9.68           &       $<$39.25           &        0.79           &           F\\
        NGC3605           &        20.7           &       9.51           &       39.12           &        0.67           &           F\\
        NGC3610           &       21.4           &       10.19           &       39.62           &        0.76           &           L\\
        NGC4239           &       17.0           &       9.25           &       $<$39.86           &        0.65           &           F\\
        NGC4342           &       16.1           &       9.31           &       $<$40.16           &        1.47           &           Fe\\
        NGC4387           &       21.4           &       9.73           &       39.97           &        0.72           &           F\\
        NGC4417           &       16.1           &       9.78           &       $<$40.69           &        0.71           &           Ra\\
        NGC4434           &       26.7           &       9.90           &       $<$40.28           &         0.70           &           F\\
        NGC4464           &       16.1          &       9.21           &       $<$39.82           &        0.88           &           F\\
        NGC4467           &       16.1           &       8.70           &       $<$39.30          &        0.98           &           F\\
        NGC4474           &       16.1           &       9.65           &       $<$39.86           &        0.72           &           Re\\
        NGC4494           &       17.0           &       10.67           &       39.91           &        0.55           &           L\\
        NGC4503           &       16.1           &       9.78           &       $<$39.89           &        0.64           &           Re\\
        NGC4550           &       15.8           &       9.72           &       39.78           &        0.89           &           Fe\\
        NGC4551           &        17.3           &       9.65           &       $<$39.16           &         0.80          &           F\\
        NGC4564           &          15.0           &       9.81           &       $<$39.80          &         0.80         &           Re\\
        NGC4621           &       18.3          &       10.44           &       40.14           &        0.85           &           L\\
        NGC4648           &       24.8           &       9.88           &       $<$39.90          &        0.92           &           Re\\
        NGC4660           &       16.1           &       9.75           &       $<$39.40          &        0.91           &           L\\
        NGC4697           &       11.7           &       10.33           &        39.90        &        0.74           &           F\\
        NGC4742           &       15.5           &       9.75           &       $<$39.99           &        1.09           &           F\\
        NGC4881           &       89.5           &       10.34           &       $<$40.33           &        0.76           &           F\\
        NGC5308           &       28.2           &       10.21           &       $<$40.02           &        0.82           &           Re\\
        NGC5812           &       26.9           &       10.27           &       $<$40.40          &        0.59           &           Re\\
        NGC5838           &       23.2           &       10.21           &       40.03           &        0.93           &           Ra\\
        NGC5845           &       26.0           &       9.67           &      $<$40.05           &        0.51           &           F\\
        NGC7332           &       23.0           &       10.21           &       $<$40.36           &         0.90           &           F\\
        NGC7457           &       13.2           &       9.96           &       $<$39.68           &        0.61           &           L\\
        NGC7743           &        20.7           &       9.93           &       39.58           &         0.50       &           Ra\\
\noalign{\smallskip}
\hline
\end{tabular} 
$^a$ Distance (see Sect. 2).\\
$^b$ Blue band luminosity (see Sect. 2).\\
$^c$ Soft X-ray luminosity for the whole galaxy (see Sect. 2).\\
\end{table}
\begin{table}
\contcaption{}
$^d$ Slope of the surface brightness profile in the central region
($I\propto R^{-\gamma}$). The references for the $\gamma $ values are
given in the last column: (L) Lauer et al. 2005; 
(Re) Rest et al. 2001; (F) Faber et al. 1997;
(Ra) Ravindranath et al. 2001; (Fe) Ferrarese et al. 1994; (Q) Quillen et
al. 2000; (C) Crane et al. 1993.
\end{table}

\begin{table}
\caption[] { Properties of galaxies with $\gamma$ and \ln measured}
\begin{tabular}{ l  l  l  l  l l  l }
\noalign{\smallskip}
\hline
\noalign{\smallskip}
 Name & d$^a$  & log \lb$^b$ & log L$_{\rm HX,nuc}^c$ & & $\gamma^{d} $ & \\
      & (Mpc)  & $(L_{B\odot})$    &(erg s$^{-1})$    &      &       &  \\
\noalign{\smallskip}
\hline
\noalign{\smallskip}
Core:\\
        NGC404            &        2.4           &       8.47           &       $<$37.52     & Liu   &     0.28           &   Ra      \\
        NGC507            &        60.3           &       10.87           &       $<$39.92     & Do   &     0.01           &   L      \\
         NGC720           &       27.7           &       10.63           &        37.90       & Je   &     0.06          &   F      \\
        NGC1052           &       19.4           &        10.20          &       41.26        & Du   &     0.11          &   Ra     \\
        NGC1316           &       21.5           &       11.08           &       39.62        & KF   &     0.13          &   L      \\
        NGC1399           &       20.0           &        10.60          &       $<$38.96     & Lo   &     0.09       &   L      \\
        NGC1600           &       60.8           &       11.04           &       $<$39.59     & Si   &     0.08       &   F      \\
        NGC2300           &       28.8           &        10.44          &       40.91       & LB   &     0.08           &           L\\
        NGC2841           &       13.0           &       10.62           &       38.33        & Du   &     0.01          &   F      \\
         NGC3193           &       34.0           &       10.55           &      $<$39.69     &RW    &     0.01           &           Re\\
        NGC3379           &       10.6           &       10.11           &       38.70         & Da   &     0.18          &   L      \\
        NGC3607           &        22.8           &       10.58           &       40.28        & Te   &     0.26          &   L      \\
        NGC3842           &       91.8           &        11.02          &       $<$39.6      & Su   &     0.15       &   L     \\
        NGC4261           &       31.6           &        10.70          &       41.05        & Do   &     0.00       &   Ra     \\
        NGC4278           &       16.1           &       10.23           &       40.08        & Du   &     0.06          &   L      \\
        NGC4291           &       26.1           &       10.05           &       39.82        & RW   &     0.02          &   L      \\
        NGC4365           &       20.4           &       10.56           &       $<$37.96     & Si03 &     0.09       &   L      \\
        NGC4374           &       18.4           &        10.70          &       39.11        & Du   &     0.13          &   Ra     \\
        NGC4382           &       18.4           &       10.77           &       $<$37.78     & Si03 &     0.01       &   L      \\
        NGC4406           &       17.1           &       10.73           &       40.28        & CM   &    -0.04          &   L      \\
        NGC4472           &       16.3           &       10.92           &       38.46        & Bil  &     0.01          &   L      \\
        NGC4476           &       17.2           &       9.50           &       $<$39.34           &  Liu &     0.21           &           Fe\\
        NGC4478           &       18.1           &       9.90           &       $<$39.15           & Liu  &    0.10           &           L\\
        NGC4486           &       16.1           &       10.86           &       40.52        & Du   &     0.25          &   F      \\
        NGC4552           &       15.4           &       10.26           &       39.33        & Fi   &    -0.02          &   L      \\
        NGC4636           &       14.7           &       10.44           &       $<$38.41     & Lo   &     0.13       &   Ra     \\
        NGC4649           &       16.8           &       10.78           &       37.82        & Sol  &     0.16          &   L      \\
        NGC4874           &       89.5           &       11.07           &       $<$39.94     & Do   &     0.13       &   F      \\
        NGC5548           &       68.9           &       10.67           &       43.36        & Bi   &     0.20          &   Ra      \\          
       NGC5813           &       32.2           &       10.68           &       $<$39.44        & Liu   &     -0.08          &   L      \\          
       NGC6166           &       110.1           &       11.21           &       42.56        & Do   &     0.08          &   F      \\          
        NGC7052           &        62.4           &        10.50           &       $<$40.48    & Do   &     0.16      &   Q      \\
       NGC7213           &       21.5           &       10.22           &       42.08       & Bi   &     0.21          &   L     \\          
        IC1459           &       29.2           &       10.75           &       40.88         & Fa   &     0.15           &   L      \\
\noalign{\smallskip}
Intermediate:\\
        NGC821           &        24.1           &       10.28           &       $<$38.36        & Fa04 &     0.42          &   L       \\
        NGC3862           &        83.1           &       10.57           &       42.32           & Do   &     0.39             &   C      \\
        NGC4594           &        9.8           &       10.74           &       40.21           & Pe   &     0.40            &   C      \\
        NGC5273           &        16.5           &       9.65       &       39.49               & LB   &     0.37            &   Ra      \\
\noalign{\smallskip}
Power law:\\
        NGC221           &       0.81            &        8.46           &       35.97           & Ho   &     0.50           &   Ra     \\
        NGC1553           &       18.5           &       10.84           &       39.81           & Bl   &     0.74           &   Q      \\
       NGC2974           &       21.5           &       10.26           &       40.32      & LB &        0.62           &           L\\
        NGC3065           &       30.4           &       9.75           &        41.32           & Iy   &     0.79           &   Re     \\
        NGC3115           &       9.7           &       10.18           &        39.5       & LB         &        0.52           &           L\\
        NGC3377           &       11.2           &        9.82           &       38.66           & RW   &     0.62           &   L      \\
       NGC3384           &       11.6           &       9.98           &       $<$38.77     &RW           &        0.71           &           L\\
        NGC4143           &       15.9           &       9.91           &       39.98            & TW   &     0.59            &   Ra     \\
        NGC4387           &       21.4           &       9.73           &        $<$39.90        & CM   &     0.72         &   F      \\
        NGC4464           &       16.1          &       9.21           &       $<$39.07           &Liu &        0.88           &           F\\
        NGC4467           &       16.1           &       8.70           &       $<$38.93          &  Liu &      0.98           &           F\\
        NGC4494           &       17.0           &       10.67           &       38.86           & Du   &     0.55           &   L      \\
        NGC4550           &       15.8           &       9.72           &      $<$38.36          & TW   &     0.89           &  Fe     \\
        NGC4564           &          15.0           &       9.81           &       38.56            & So   &       0.80          &    Re     \\
        NGC4697           &       11.7           &       10.33           &       38.44           & Sa   &     0.74           &   F      \\
        NGC5845           &       26.0           &       9.67           &       39.15            & So   &      0.51           &    F      \\
        NGC7332           &       23.0           &       10.21           &       $<$39.70           &  Liu    &   0.90           &           F\\
        NGC7743           &        20.7           &       9.93           &       39.43            & Te   &       0.50          &    Ra     \\
\noalign{\smallskip}
\hline
\end{tabular} 

$^a$ Distance, as for Tab. 1.
\end{table}

\begin{table}
\contcaption{}

$^b$ Blue band luminosity for the whole galaxy, as for Tab. 1.

$^c$ Nuclear X-ray luminosity in the 2--10 keV band.
References for the X-ray luminosity values are given in the next column:
(Bi) Bianchi et al. (2004);
(Bil) Biller et al. 2004;
(Bl) Blanton et al. 2001; 
(CM) Colbert \& Mushotzky (1999); 
(Da) David et al. (2005);
(Do) Donato et al. 2004; 
(Du) Dudik et al. 2005; 
(Fa) Fabbiano et al. 2003;
(Fa04) Fabbiano et al. 2004;
(Fi) Filho et al. 2004; 
(Ho) Ho et al. 2003; 
(Iy) Iyomoto et al. 1998;
(Je) Jeltema et al. 2003;
(KF) Kim \& Fabbiano 2003; 
(LB) Liu \& Bregman (2005);
(Liu) Ji-Feng Liu (private comm., based on LB);
(Lo) Loewenstein et al. 2001; 
(Pe) Pellegrini et al. 2003; 
(RW) Roberts \& Warwick 2000; 
(Sa) Sarazin et al. 2001;
(Si) Sivakoff et al. 2004; 
(Si03) Sivakoff et al. 2003;
(So) Soria et al. 2005, submitted to ApJ;
(Sol) Soldatenkov et al. 2003;
(Su) Sun et al. (2005); 
(Te) Terashima, Ho \& Ptak 2000; 
(TW) Terashima \& Wilson 2004.

$^d$ Slope of the central profile, and references for the $\gamma $ values in
the next column, as for Tab. 1.
\end{table}

\section*{Acknowledgments}
I warmly thank Ji-Feng Liu for providing the upper limits on
the flux for 7 of the objects listed in Table 2 and the referee for useful
comments.
I acknowledge use of the NASA Extragalactic database (NED), operated by the
Jet Propulsion Laboratory, California Institute of Technology, and of the
LEDA database.

\end{document}